\documentclass[a4paper,fleqn,usenatbib]{mnras}

\usepackage{newtxtext,newtxmath}

\usepackage[T1]{fontenc}
\usepackage{ae,aecompl}

\usepackage{graphicx}	
\usepackage{subfigure}
\usepackage{amsmath}	
\usepackage{amssymb}	

\title[Merger delay time distribution of EE sGRBs]{Merger delay time distribution of extended emission short GRBs}

\author[Anand et al.]{
Nikhil Anand \thanks{E-mail: nikhil.31.anand@gmail.com},
Mustafa Shahid, 
and Lekshmi Resmi 
\\
Indian Institute of Space Science and Technology, Trivandrum, India.
}

\date{Accepted XXX. Received YYY; in original form ZZZ}

\pubyear{2017}

\begin{document}
\label{firstpage}
\pagerange{\pageref{firstpage}--\pageref{lastpage}}
\maketitle

\begin{abstract}
The most popular progenitor model for short duration Gamma-Ray bursts (sGRBs) is the merger of two compact objects. However, the short GRB population exhibit a certain diversity: some bursts display an \textit{extended emission} (EE), continuing in soft $\gamma$-rays  for a few hundreds of seconds post the initial short pulse. It is currently unclear whether the origin of such bursts is linked to compact object mergers.

Within the merger hypothesis, the redshift ($z$) distribution of short GRBs is influenced by the \textit{merger delay time}, i.e., time elapsed between the merger and the formation of the binary star system, which is dominated by the time-scale for gravitational wave losses during the compact binary phase. We examine redshift distributions of short GRBs with extended emission to see whether their formation channel requires considerable delay post the star formation episode. Our results show that the $z$ distribution of EE bursts is consistent with the merger model. We attempted to compare the delay time distribution of the EE and the non-EE short bursts. However, no statistically significant difference could be seen within the limited sample size.
\end{abstract}




\section{Introduction}
The bimodal distribution of Gamma Ray Bursts (GRBs) in their temporal-spectral plane, as long-soft and short-hard bursts \citep{Kouveliotou:1993yx}, calls for a distinction in their progenitor models. The most popular model of short duration Gamma Ray Bursts is the merger of compact binaries under loss of orbital energy and angular momentum through Gravitational Waves (GWs).  Leading merger models of short bursts involve double neutron star (DNS) and Neutron Star Black-Hole (NSBH) systems \citep{Paczynski:1986, Eichler:1989}.

A direct confirmation of the merger hypothesis has to wait till a joint detection by gravitational wave and electromagnetic detectors. However, several indirect observational characteristics of short bursts are favourable to this hypothesis. Long GRBs show a conclusive association with core-collapse supernovae \citep{Galama:1998ea, Hjorth:2003}, while short bursts do not. Positional offsets from the photo center of host galaxies are systematically larger than that of long bursts or core collapse supernovae, indicating natal kick velocities typical of neutron stars that can propel the binary off its birth place \citep{Fong:2013iia}. sGRB afterglows, being systematically fainter than that of long ones \citep{Fong:2015oha}, may demand a lower density ambient medium typical of high galactic latitudes where compact binaries are expected to reach due to natal kicks. Short GRB hosts are of diverse types, implying a broad distribution in the time elapsed since the star formation epoch \citep{Fong:2013eqa}. Finally, recent claims of infra-red excess in short GRB afterglows is a promising signature of merger novae, arising from tidally disrupted neutron rich material from the merger \citep{li1999radio, Tanvir:2013pia}. 

Under the merger hypothesis, the redshift ($z$) distribution of sGRBs is a convolution of the star formation history of the universe and the merger delay time, where the latter is the time elapsed between the formation of a stellar binary and its eventual merger as two compact objects. The delay time is divided into two phases. First, the time taken for the two stars to evolve and become compact objects, which is typically of the order of $10^6$ yrs. A major fraction of the delay time, $\sim 10^9$yrs \citep{HulseTaylor:1975, Postnov:2014tza}, is spent in the second phase, as two compact objects spiralling in by the emission of gravitational waves. Hence, merger delay time is an indication of the gravitational wave loss time-scale, and the $z$-distribution will be carrying an imprint of this. 
%


As mentioned earlier, there are two progenitor channels, DNS and NS-BH systems, typically discussed in the context of short GRBs. Even within the DNS scenario, the central engine of the burst, formed as the result of the merger, can be diverse: a BH-torus system, a hypermassive short lived NS, or a long-lived NS, depending on the mass and equation of state of the components. Hence, it is likely that there could be a diversity within the short bursts themselves, and indeed some observations indicate the existence of such.

A unique characteristic shown by some short bursts is the presence of soft $\gamma$-ray emission ($2 - 25$~keV) extending to several hundreds of seconds post the initial short hard pulse \citep{Villasenor:2005}.  In some cases, the energy emitted in the extended emission is larger, even by a factor of $30$ \citep{Perley:2008ay}, than that in the prompt spike. Around $25$\% of short bursts are associated with extended emission (EE) \citep{Norris:2009rg}. This long lasting prompt emission challenges models invoking a BH-torus system at the end of the merger. One of the alternate propositions is magnetar central engines, forming from DNS binaries or from accretion induced collapse of white dwarf binaries \citep{Metzger:2008}. 

The absence of spectral lag in the prompt emission, a characteristic differentiating short bursts from their long duration cousins, is shared by both EE and non-EE bursts \citep{Norris:2006rw}. However, there is no conclusive evidence for any defining characteristics the EE population may have from the remaining sGRB population. Though \cite{Troja:2007kt}  claimed that EE bursts potentially have a smaller host offset compared to other short bursts, studies based on a larger sample found that there are no conclusive differences in the offsets or host types of EE bursts from those of other short bursts \citep{Fong:2013iia, Fong:2013eqa}. 

It is possible that EE-sGRBs are a distinct class which originates from a separate type of merging compact binaries compared to the rest of the short bursts. Or, the EE component could be an indication to a different kind of central engine.  %
Since the timescale for merger under gravitational wave radiation losses depends on the component masses, eccentricity, and orbital period of the binary (see eq-3 in \cite{Postnov:2014tza}), naively one can expect different merger time distributions for different classes of merging binaries. 

In this paper, we infer the delay time distribution of EE short GRBs from their redshift distribution under the ambit of the merger model. If EE short GRBs have  a massive star origin similar to long GRBs, we expect to see its reflection in the inferred delay time distribution. Further, we compare the delay time of EE bursts with that of the non-EE short GRBs, albeit the caveat that the relative faintness of EE emission may lead to non-detections in higher redshifts. If a distinction in progenitor channel exists for the EE bursts, an imprint of the delay time would be visible in the redshift distribution in comparison with other short bursts. In section-\ref{sect2}, we describe the sample selection. Theoretical modelling of the redshift distribution is explained in section-\ref{sect3}. In section-\ref{sect4} we describe the results of our analysis. We summarize the paper in section-\ref{sect5}.

\section{Sample of short GRBs}
\label{sect2}
We first constructed a sample of short duration GRBs detected by \emph{Swift} with reliable redshift measurements. To keep the sample uniform, we did not consider bursts detected by instruments other than \emph{Swift} (for example, GRB 050709 detected by \emph{HETE} is not part of the sample). This  allows us to simplify the problem without having to consider varied biases across instruments. Since nearly all redshift measurements are available for \emph{Swift} bursts alone, restricting to \emph{Swift} does not compromise much on the sample size. 

We used the 3rd \emph{Swift} BAT catalogue \citep{Lien:2015oqa} and the compilation by \citep{Fong:2015oha} to construct the sample. Both authors use $T_{90} < 2$~sec criteria to categorize a burst as short. Our full GRB list contains only bursts up to 2015 October, where the BAT catalogue ends \footnote{During the preparation of this manuscript we came across \citep{Gibson:2017dep} which includes GRB160410A, an EE burst at $z = 1.7$, the largest $z$ for an EE burst till date. We ran separate simulations after including this burst.}. Redshift measurements are available only till March 2015 in \cite{Fong:2015oha}. For the one burst in our sample beyond this period, GRB150423A we used the BAT catalogue and the online table of Jochen Greiner\footnote{http://www.mpe.mpg.de/~jcg/grbgen.html.} for $z$ measurement. For GRB 080123, we use the $z$ value given in \citep{Leibler:2010uq}, since \citep{Fong:2016orv} does not list this burst. 

We excluded bursts with confusing redshift information (for example, 050813 and 051210) from the sample. GRBs 060505 and 090927, displaying spectral lag or hardness ratio typical of long bursts are not included. The SN-less $100$~s duration GRB 060614 is also excluded. Two other bursts, 070506 and 090530, which are debated to be short bursts but having a main pulse longer than $2$~s are also not part of the sample. 
%
%
\subsection{EE sample}
\citep{Lien:2015oqa} list short bursts with \textit{definite} (table-3 of original paper) and \textit{possible} (table-4 of original paper) extended emission. This sample (Lien-sample for the rest of the paper),  is a collection of all previous literature including GCN circulars on EE searches. Possible shorts are the ones where the pulse is slightly longer than $2$~s, or when the EE is weak, or if its existence is confusing due to stronger sources in the field of view. 

We cross checked the list with the $T_{90} < 2$~s bursts in five authors searching extended emission in GRBs \citep{Gompertz:2013zga, DAvanzo:2014urr, Kagawa:2015gaa, Kaneko:2015lga, Fong:2016orv}. The sample selection criteria used by various authors are different, leading to a certain level of disagreement in results. 

All $T_{90} <2$~s bursts for which \cite{Gompertz:2013zga} detects EE are classified as EE in \cite{Lien:2015oqa}. Their sample extends only till 2011. All $T_{90} <2$~s bursts of the D\' Avanzo sample are included in the Lien-sample. Since they exclude bursts at  low Galactic extinction lines of sight, their sample is much smaller than that of Lien. Kagawa finds three short bursts with EE which are also part of the Lien-sample.  Eight of Kaneko's EE  bursts have $z$ information, of which the five with $T_{90} < 2$~s are part of the Lien-sample. Duration of two bursts (051016B, 070506) are larger than $2$~sec and the classification of 090927 is dubious. Hence, their absence in Lien-sample is justified.

GRB 061006 and 061210 are absent from the sample of  Kagawa and D\' Avanzo due to their selection criteria  

However, \cite{Lien:2015oqa} report EE in some short bursts which other authors failed to detect. \cite{Kagawa:2015gaa} and \cite{DAvanzo:2014urr} considered GRB 090510 but did not detect EE. %
\cite{DAvanzo:2014urr} reports a failed EE detection in GRB 100816A, while it is included as a possible EE burst in \cite{Lien:2015oqa} because the main pulse is longer than $2$~sec. Because of the pulse duration we have excluded it from our sample. 

Hence, we see that none of the conflicts between the authors are relevant here, and proceed to use \cite{Lien:2015oqa} as the reference for EE bursts. Our definite-EE sample has all Lien EE-bursts with confirmed redshift measurements (excluded the ones having only upper or lower limits of $z$), which is $5$ bursts. For the full EE sample, we add two weak EE detections from Lien's table-4.  Other bursts classified as possible EE in their table-4 are not considered because they have longer durations or the EE detection is confusing. In table-1, we lists all the bursts in our sample. 

\begin{table}
	\centering
	\label{tab:firsttable}
\begin{tabular}{lcc}
\hline
\multicolumn{1}{c}{\textbf{GRB}} & \textbf{z}                & \textbf{Sample} \\ \hline
050724                           & \multicolumn{1}{l}{0.257} & EE              \\
061006                           & 0.4377                    & EE              \\
061210                           & 0.41                      & EE              \\
070714B                          & 0.923                     & EE              \\
071227                           & 0.381                     & EE              \\
080123                           & 0.495                     & weak EE         \\
090510A                          & 0.930                     & weak EE         \\\hline 
050509B                          & \multicolumn{1}{l}{0.225} &                 \\
051221A                          & 0.546                     &                 \\
060502B                          & 0.287                     &                 \\
060801                           & 1.1304                    &                 \\
061201                           & 0.111                     &                 \\
061217                           & 0.827                     &                 \\
070429B                          & 0.902                     &                 \\
070724A                          & 0.457                     &                 \\
070729                           & 0.8                       &                 \\
070809                           & 0.473                     &                 \\
080905                           & 0.122                     &                 \\
090426A                          & 2.609                     &                 \\
090515A                          & 0.403                     &                 \\
100117A                          & 0.915                     &                 \\
100206A                          & 0.407                     &                 \\
100625A                          & 0.452                     &                 \\
101219A                          & 0.718                     &                 \\
111117A                          & 1.3                       &                 \\
120804A                          & 1.3                       &                 \\
130603B                          & 0.356                     &                 \\
131004A                          & 0.717                     &                 \\
140622A                          & 0.959                     &                 \\
140903A                          & 0.351                     &                 \\
141212A                          & 0.596                     &                 \\
150101B                          & 0.134                     &                 \\
150120A                          & 0.46                      &                 \\
150423A                          & 1.394                     &                 \\  \hline \hline
\end{tabular}
\caption{The sample of short GRBs with redshift information, primarily based on \citep{Lien:2015oqa} and \citep{Fong:2015oha}. We have also considered several other literature on sGRB EE emission before arriving at the EE sample. See text for details. The last column indicates whether a burst is part of the EE sample or not.  }
\end{table}

\begin{figure}
\label{sample}
\includegraphics[scale=0.4]{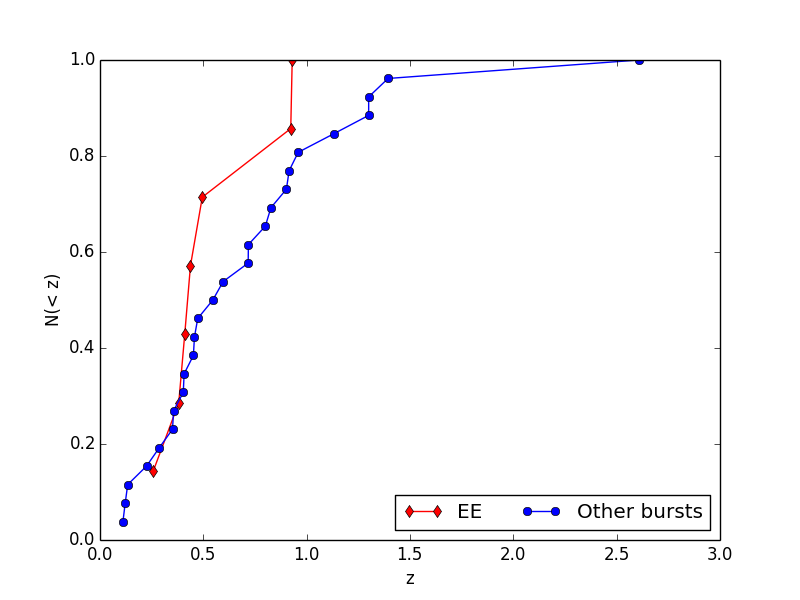}
\caption{The cumulative $z$ distribution of EE (red) bursts in our sample, along with the non-EE (blue) bursts (see table-1). The lower redshifts of EE bursts could be due to the apparent faintness of the EE emission.}
\end{figure}

Figure-1 is the cumulative $z$ distribution of the bursts. It is obvious from the distribution that the EE bursts are confined to smaller redshifts. This could potentially come from a selection bias if EE bursts are fainter on an average, hampering their $z$ determination. We have not included GRB160410A in this figure, which we include in a final special run of simulations alone. 



We proceed to model the cumulative $z$ distribution of the EE bursts under the merger hypothesis to see the most agreeable merger delay time distribution. We also do a comparison between the EE and the non-EE sample. But the small size of the EE sample can affect the statistical significance of the result while comparing the EE and non-EE sample. Moreover a potential selection bias due to the faintness of the EE emission can interfere in the non-EE sample (if the EE is too faint to be detected, the burst will get classified as a non-EE burst) \citep{Perley:2008ay, Norris:2009rg}. 

\section{Modelling redshift distribution}
\label{sect3}
The broad framework we adopt to model the redshift distribution is well established in the literature \citep{Guetta:2004fc, Nakar:2005bs, Petrillo:2012ij, Hao:2013mba, Wanderman:2014eza}. Nevertheless, for the sake of completion, we describe it below.

Rate of compact object mergers per unit co-moving volume, $\rho_{\rm merg}$, is a convolution of the star formation history of the universe $\rho_{\rm SFR} (z)$ and delay time distribution $f(\tau)$ representing the probability that the merger happens with a time delay $\tau$.
\begin{equation}
\rho_{\rm merg} (z) = A \int_z^\infty \rho_{\rm SFR} (z^\prime) f(\tau(z^\prime)) \frac{dt}{dz^\prime} dz^\prime.
\label{rhomerg}
\end{equation}

$\tau$, the time elapsed between two epochs represented by $z$ and $z^\prime$,  can be written as $(t(z) - t(z^\prime)$ where $t(z)$ is the age of the universe at redshift $z$. The coefficient $A$ here takes care of the fraction of stars ending up as the kind of compact binaries in consideration. This fraction is assumed to be independent of redshift, and hence will not enter the normalized expression. 
  
If all mergers lead to short GRBs, the rate $\dot{N}_{\rm SHB}$ of short hard bursts per unit $z$ per unit time detected by an ideal detector is given by,
\begin{equation}
\frac{d\dot{N}_{\rm SHB}}{dz} = \frac{\rho_{\rm merg}}{(1+z)} \frac{dV}{dz} 
\end{equation}

However, for a detector with a limiting flux sensitivity of $f_{\rm lim}$ which will not be able to detect bursts of all luminosity, the number of detectable bursts will depend on the luminosity function $\phi(L)$.  Hence, 
\begin{equation}
\frac{d\dot{N}_{\rm obs}}{dz}  = \int_{L_{\rm min}}^{L_{\rm max}} dL \;  \phi(L)  \frac{d\dot{N}_{\rm SHB}}{dz}
\end{equation}
where $L_{\rm min}$ and $L_{\rm max}$ are the lowest and highest luminosities in the burst distribution. If $L_{\rm min} < L_{\rm lim}$, the minimum detectable luminosity corresponding to $f_{\rm lim}$, $L_{\rm min}$ will be replaced with $L_{\rm lim}$, which is a function of distance and hence $z$.

Hence, the normalized cumulative distribution at a redshift $z_0$ can be written as, 
\begin{equation}
{\cal N}(<z_0) \propto \int_{0}^{z_0} dz \;  \int_{L_{\rm min}(z)}^{L_{\rm max}} dL \;  \phi(L) \frac{d\dot{N}_{\rm SHB}}{dz}
\label{cumulN}
\end{equation}

Thus, the major components of the model are the star formation history ($\rho_{\rm SFR}(z)$), delay time distribution $f(\tau)$, and the luminosity function $\phi(L)$ of short GRBs. While there is a reasonable understanding of the cosmic star formation history upto a moderate redshift \citep{Madau:2014bja}, the merger delay time distribution and short GRB luminosity function are not well understood. For both these components we will have to resort to empirical prescriptions.


\begin{figure*}
\begin{center}
\includegraphics[scale=0.3]{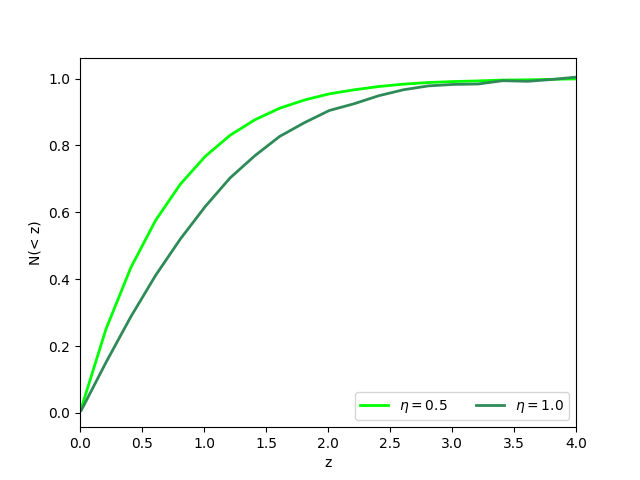}
\includegraphics[scale=0.3]{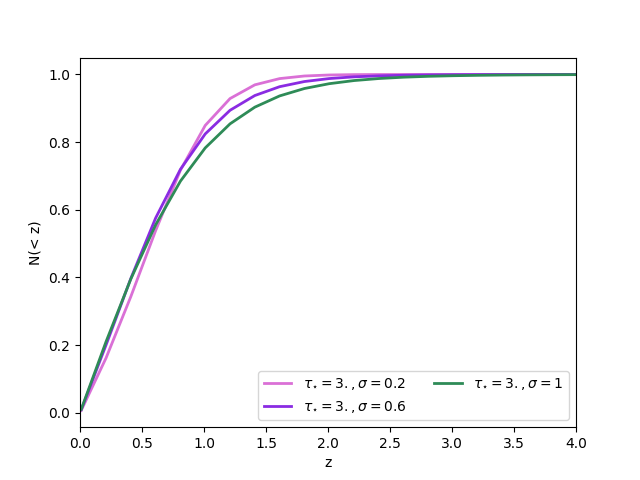}
\includegraphics[scale=0.3]{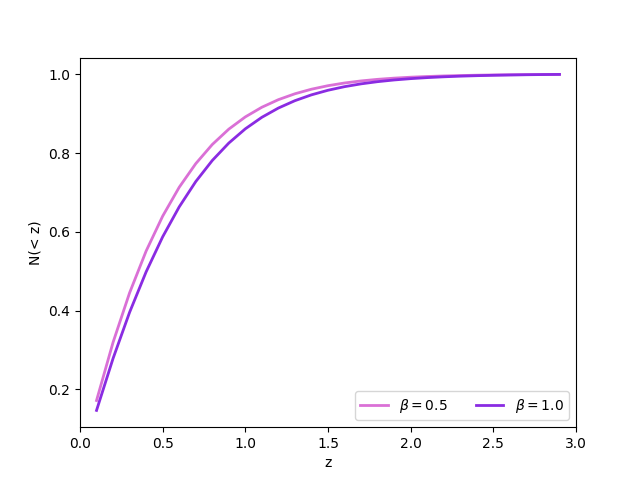}
\caption{Model predictions of the cumulative distribution ${\cal{N}}(<z_0)$. Dependencies of the model on parameters of the empirical $f(\tau)$ is shown. Left: PL model. A flat power-law $f(\tau)$, i.e.,  smaller values of $\eta$, implies an increased probability for larger delay times.  Hence, the model curve corresponding to the flatter delay time ($\eta =0.5$) saturates before $\eta = 1$ indicating most of the merger occuring at lower redshifts. Middle: LN model. No significant change in the final distribution due to variation in $\sigma$. Right: effect of variation in luminosity function index.}
\label{trend}
\end{center}
\end{figure*}

\subsection{Components of the model}
\emph{Star formation history of the universe:} There is considerable uncertainity in the star formation history beyond $z \sim 2$. \cite{Madau:2014bja} uses a set of post-2006 galaxy surveys in rest-frame Far-UV, Mid, and Far IR to extract a ready-to-use functional form of the cosmic star formation history. The survey data they use is restricted to $z  < 8$. The best fitting function is given by,
\begin{equation}
\rho_{\rm SFR}(z) = 0.015\frac{(1+z)^{2.7}}{1 + [(1+z)/{2.9}]^{5.6
}} M_{\odot} \; {\rm year}^{-1} \; {\rm Mpc}^{-3}.
\end{equation}
This function rises in the early universe (between $3 \lesssim z \lesssim 8$), peaks at a redshift of around $1.5$ -- $2$, and gradually declines to the present day universe. It shows a steady behaviour in the high-z range.  There is some debate in the literature about the nature of the SFR for $z > 9$ (see \cite{Madau:2014bja} for details). Though such large redshifts are important to our analysis, since there is no definitive answer to the nature of high-$z$ SFR, we resort to the \citep{Madau:2014bja} prescription which uses the most recent state-of-the-art surveys. 

\emph{Delay time distributions:} In the recipe for delay time, we are ignoring the time taken for the two main sequence stars to evolve into compact objects. The general expression for gravitational wave loss time scale for a compact binary depends on the period,  eccentricity, the reduced mass, and the total mass of the binary system \citep{Postnov:2014tza}.

The simplest empirical form of delay time distribution $f(\tau)$, is a power-law function in $\tau$. If the initial orbital separation $a_i$ of the binaries have a power-law distribution of the form $f(a_i) \propto a_i^{q}$, the gravitational wave delay time will be distributed as $f(\tau) \propto \tau^{(q-3)/4}$ \citep{Piran1992}. This calculation ignores any possible spread in orbital eccentricity. Assuming $q = -1$ in accordance with the X-ray binary data in our galaxy, \cite{Piran1992} arrives at $f(\tau) \propto \tau^{-1}$. On the other end, there have been detailed population synthesis calculations of delay time distributions (for example, \cite{OShaughnessy:2007brt} and references therein for DNS and NS-BH systems. These calculations result in complex, at times bimodal, distributions of delay times \cite{OShaughnessy:2007brt}, which are beyond the scope of this paper to implement. 

Previous studies of short GRB rates in the literature have used simple empirical delay time distributions, mostly power-law (PL) and  log-normal (LN) functions in $\tau$ \citep{Nakar:2005bs}. The PL function is characterised by index $\eta$, while the LN function depends on two parameters, the mean delay time $\tau_\star$ and the standard deviation $\sigma$. The LN model is an approximation of a constant delay time distribution.  We also use these two empirical delay time functions in our calculations. We have used a lower cut-off of $20$~Myr in the PL model, which represents the typical timescale for single star evolution. At a given $z$, we use the age of the universe for the corresponding $z$ as an upper limit to $\tau$.  In figure-\ref{trend} we show the sensitivity of the model ${\cal{N}}(< z_0)$ on $\eta$, $\tau_\star$, and $\sigma$. We use inferences from these dependences to constrain the parameter space (see next section).

\emph{Short GRB luminosity function:} Luminosity function of short duration GRBs is convolved always with the rate, and manifests in the $z$ distribution as well as in the peak flux distribution. \cite{Nakar:2005bs} have used a single power-law luminosity function model, $\Phi(L) \propto L^{-\beta}$, since the sample size was small during that era. They found the luminosity distribution is consistent with a $\beta = 1.0$. \cite{Wanderman:2014eza} found that a double power-law model is better fitted, with an index similar to \citep{Nakar:2005bs} at the lower end of the luminosity function. Since the sample of EE bursts are small, we chose to use the simplest form of the luminosity function, a single power-law throughout this work. %

The minimum detectable luminosity at a distance $d_L(z)$ is,
\begin{equation}
L_{\rm lim} = 4 \pi d_L^2(z) k(z) f_{\rm lim}
\label{eqlmin}
\end{equation}
where $k(z)$ is the cosmological k-correction which depends on the spectral shape of the burst. This factor appears in the equation since for a burst at redshift $z$, the \emph{BAT}  energy range ($E_1 - E_2$) corresponds to emitted photons of energy $E_1 (1+z) - E_2 (1+z)$. Hence, the BAT flux of a burst with a cosmic rest-frame luminosity $L$ will be proportional to $(1+z)^{1-\alpha}$ where $\alpha$ is the spectral index ($f_\nu \propto \nu^{-\alpha}$). Most \textit{Swift} bursts have an $\alpha \sim 1$ if fitted by single powerlaw spectral models \citep{Lien:2015oqa}. Hence we can safely ignore the $(1+z)$ factor in the equation-\ref{eqlmin}. We use $f_{\rm lim} = 5 \times 10^{-9}$erg/s/cm$^{2}$ for the BAT threshold flux following \cite{Cao:2011ar}. For the range of $z$ we encounter in this analysis, $L_{\rm lim}$ varies from $10^{45}$ to $10^{51}$ and we always considered $L_{\rm min}$ to be $L_{\rm lim}$.

We restricted our analysis to $\beta > 1$ where $L_{\rm max}$ remains irrelevant. Thus, eq-\ref{cumulN} reduces to,
\begin{equation}
{\cal N}(<z_0) \propto \int_{0}^{z_0} dz \;  {L_{\rm min}}^{1-\beta}(z) \frac{d\dot{N}_{\rm SHB}}{dz}, \; \; \beta > 1.
\label{cumulN1}
\end{equation}

%
%
Throughout this paper, we have used a standard cosmological model with $H_0 = 69$~km/s/Mpc, $\Omega_m = 0.29$, $\Omega_\lambda = 0.71$.

\section{Analysis and results}
\label{sect4}

Though the star formation rate drops down towards higher redshifts in our SFR model, we checked the convergence of the integration in eq-\ref{rhomerg}. Under the assumption that the fitting function of Madau \& Dickinson is rightly reproducing the unknown SFR beyond $z$ of $8$, we performed the integration for different values of $z_{\rm max}$ and found that an $z_{\rm max}\sim 50$ ensures convergence. 
We tested the model for ($\eta > 10$) indices of the PL delay time function and found that it is able to reduce the resulting $\rho_{\rm merg}$ to the SFR. 

For a given SFR, the parameter space reduces to the two dimensional ($\beta$, $\eta$) space for the PL model. We varied the power-law index $\eta$ from $-3.$ to $0.5$. Since the variation in $\sigma$ leads to a relatively lesser deviation between the models (see figure-\ref{trend}), we fixed $\sigma$ at $0.2$ Gyr and essentially reduced the parameters space again to the two dimensional ($\beta$, $\tau_\star$) for the LN model. We increased $\sigma$ to $0.6$~Gyr but did not find any significant difference in ${\cal{N}} (<z_0)$. $\tau_\star $ was made to vary from $2.5$ to $4$~Gyr, since a mean delay above $4$~Gyr would have led to all bursts happening $z < 1$, which clearly is ruled out by the data. The index of the luminosity function $\beta$ was varied from $1.1$ to $3.$.

We constructed normalized cumulative distributions of both the data and the model. We normalized the model such that ${\cal{N}}(< z_0) = 1$ at the highest redshift of the sample. 

In order to infer the best fit delay time distribution, we undertook both $\chi^2$ analysis and the Kolmogorov-Smrinov (KS) test. Since the sample is homogeneous, we did not go for a maximum likelihood estimate. 

For the $\chi^{2}$ analysis, we computed the summed square of the deviation between the data and the model. Both the tests were done on a grid of the parameter space. 

There is a consistency between the KS and the $\chi^2$ test results. The best fit values of the parameters from the $\chi^2$ test is within the region of KS probability $p > 0.05$. Moreover, the region where $\chi^2/{\rm dof} \sim 1$ is well in agreement with the region of  $p > 0.05$ in all the runs.
\subsection{Delay time inferred for EE bursts}

We find that the redshift distribution of the EE short GRBs is consistent within the predictions of the merger models (see figures 3-10). The EE bursts seem to originate after a considerable delay from the star formation episode. Under the log-normal delay time distribution model, a mean delay of $\sim 4$~Gyr is required to produce the EE $z$ distribution. Using a power-law model for the delay time distribution, the EE bursts seem to favour a flat to positive power-law index which indicates a major fraction with increased merger delay time (than a steep negative powerlaw). However, this could be due to the faintness of EE component which restricts their detection to larger redshifts. A distribution that stops at smaller redshifts could either be due to a longer delay time or due to faint emission. We included the highest $z$ EE burst known to date, GRB160410A, to see if that can significantly reduce the delay time, but only for the PL model have we seen a change in model parameters. In PL model, as expected a trend towards a negative $\eta$ is seen, which indicates that we may get more refined results with a larger sample of EE bursts.

As expected, the non-EE bursts in comparison are consistent with a lesser delay time. But again, the statistical significance of this result is not very high due to the potential selection bias. At high $z$ values, EE need not be detected. 

Our best fit luminosity function indices are well in agreement with that of the previous studies \citep{Nakar:2005bs}. For both EE and non-EE samples, a $\beta \sim -2$ is found to be consistent, which may be an indication of a single formation channel for both EE and non-EE bursts. 

The major caveat of the analysis is the small sample size of EE bursts, especially them being restricted to relatively lower redshifts. The same can also contaminate the non-EE sample as some bursts might as well have a faint EE. Within this constraint, we infer that a \textit{delay time} of the order of Gyrs is required in the formation of these bursts.

Since the sample of EE bursts with $z$ information is small, we restricted our analysis to a  preliminary form with simplified model functions. It is possible that the underlying luminosity function is more complex. A more robust analysis with more complicated model functions can be performed once the sample size of EE bursts increases.

\section{Summary}
\label{sect5}
We model the redshift distribution of extended emission short GRBs under the ambit of the binary merger model to see the possible delay in the formation channel of this class of bursts. We find that the EE bursts require to have a considerable delay after the star formation episode in their formation channel indicating their origin in old stellar populations like compact stars. Luminosity function of both EE and non-EE bursts are consistent with each other, indicating their potential common origin. Current sample of short GRBs with $z$, especially of EE bursts are not sufficient to make a comparison of the delay time of bursts with and without extended emission. In future with a larger sample size, it will be possible to investigate their delay times separately.  

%
%
%
%
%

%
%
%
\begin{figure*}
\label{firstfig}
\begin{center}
\includegraphics[scale=0.4]{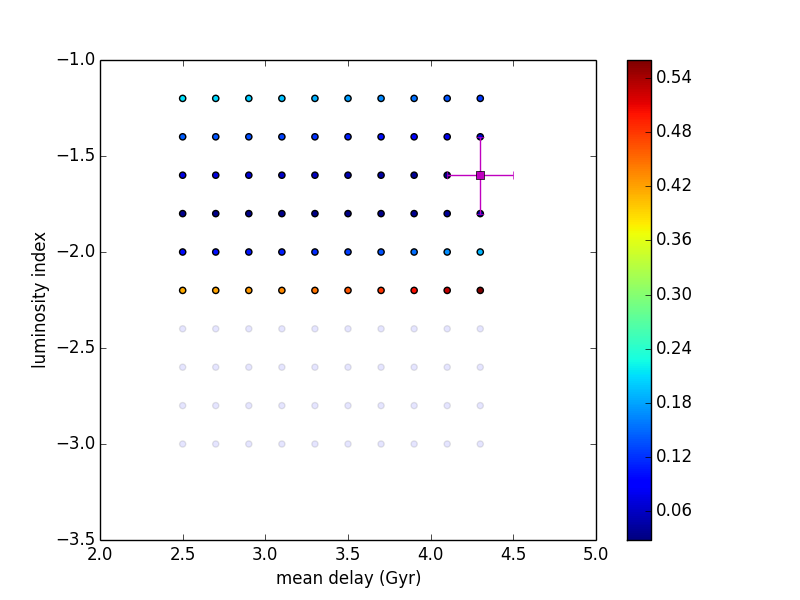}
\includegraphics[scale=0.4]{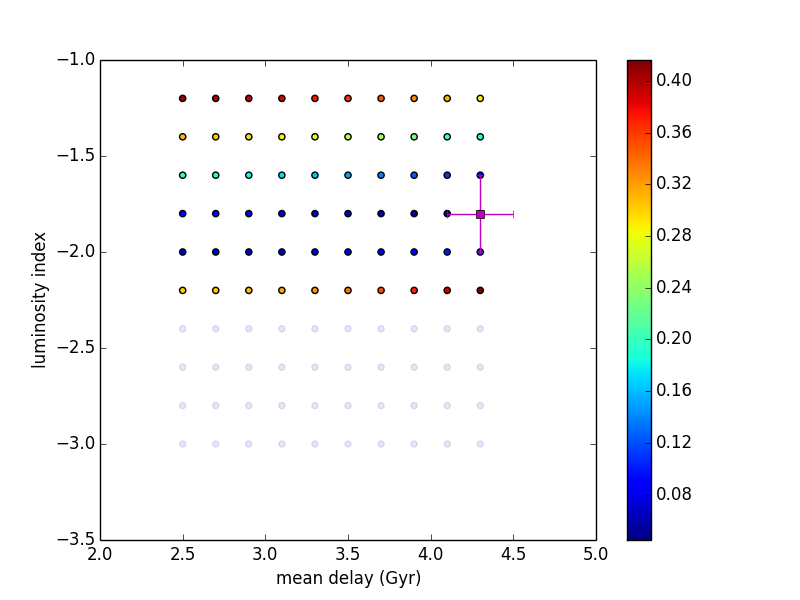}
\caption{$\chi^2$ values on grids for the lognormal model parameters ($\beta, \tau_\star$). The colored region is where $\chi^2 \sim DOF$. Figure on the left is for all EE bursts, while the figure on the right excludes the two bursts with weak EE detections.}
\end{center}
\end{figure*}
\begin{figure*} 
\begin{center}
\includegraphics[scale=0.4]{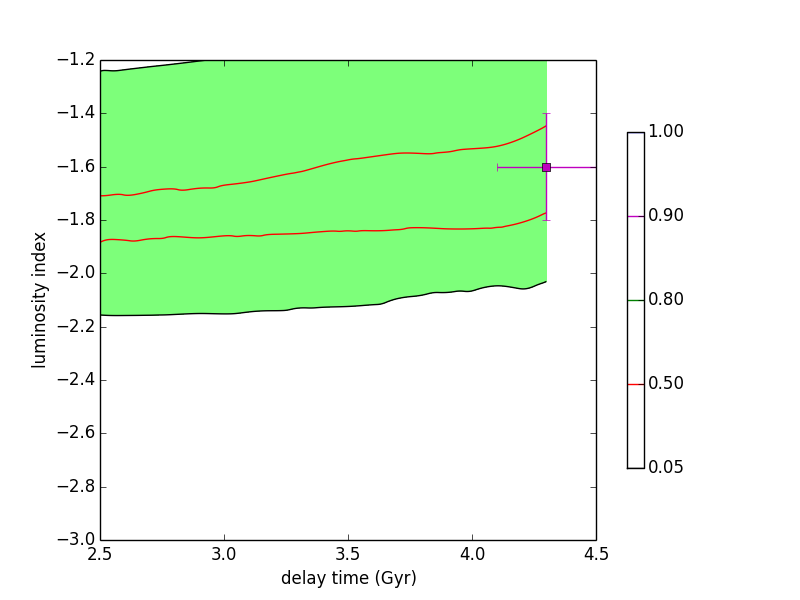}
\includegraphics[scale=0.4]{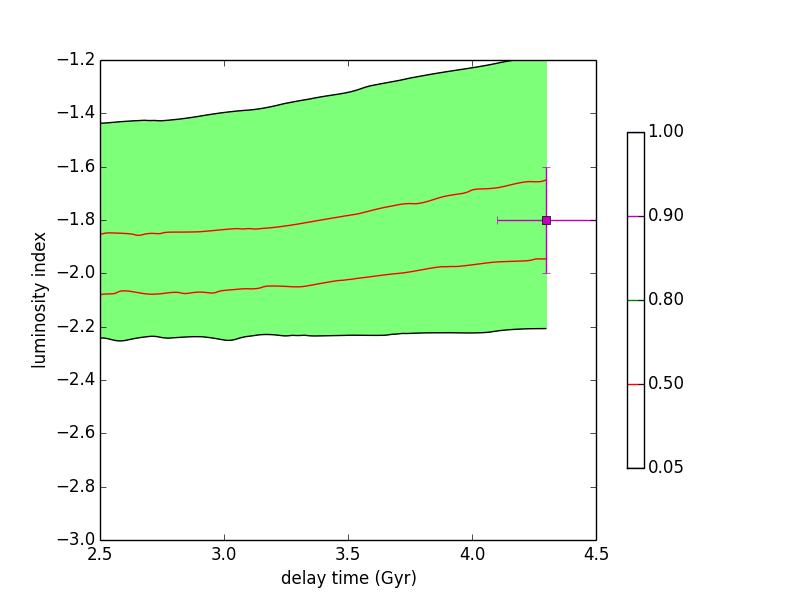}
\caption{Comparison between $\chi^2$ and KS tests for the LN model. Even though the best fit point is not coinciding with regions of higher KS probability, it is within the region where the KS probability $P > 0.05$. Left figure correspond to the full EE sample and the right one correspond to only the 5 strong detections.}
\end{center}
\end{figure*}
\begin{figure*} 
\begin{center}
\includegraphics[scale=0.5]{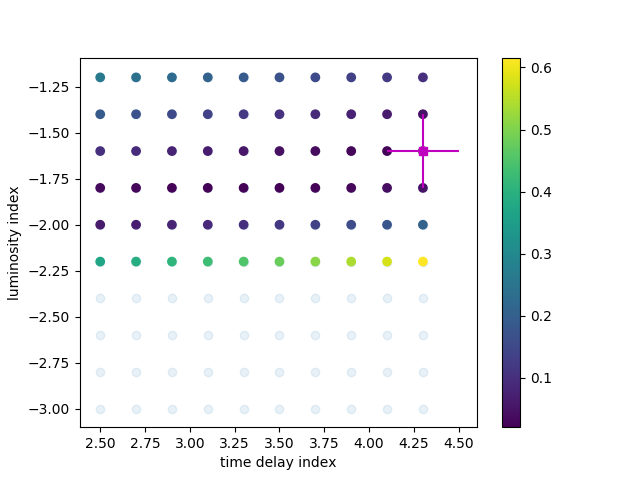}
\includegraphics[scale=0.5]{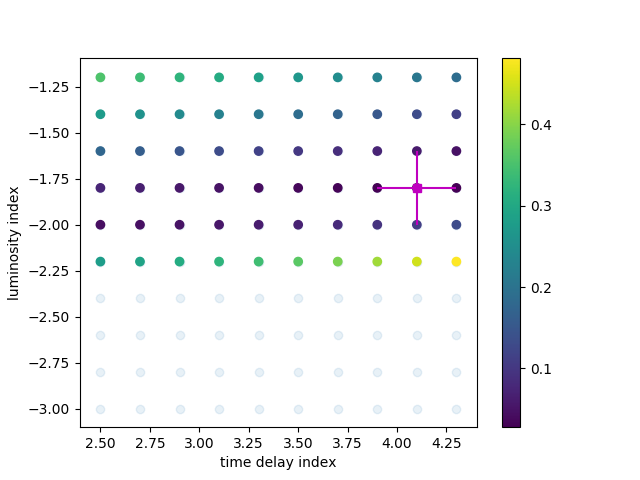}
\caption{Results of the LN model run after including 160410A, the highest redshift EE burst detected so far. No change in the resultant parameter space within the size of the grid we used for $\chi^2$.}
\end{center}
\end{figure*}
\begin{figure*}
\centering
\includegraphics[scale=0.4]{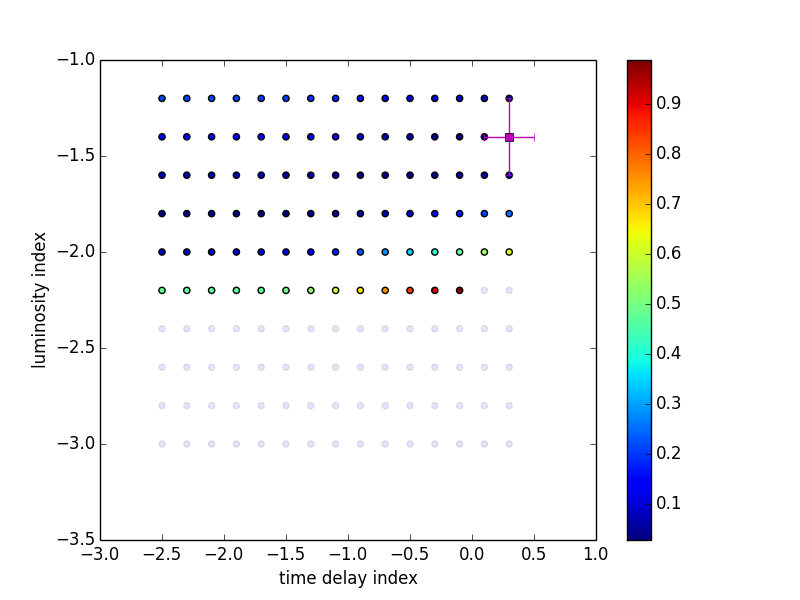}
\includegraphics[scale=0.4]{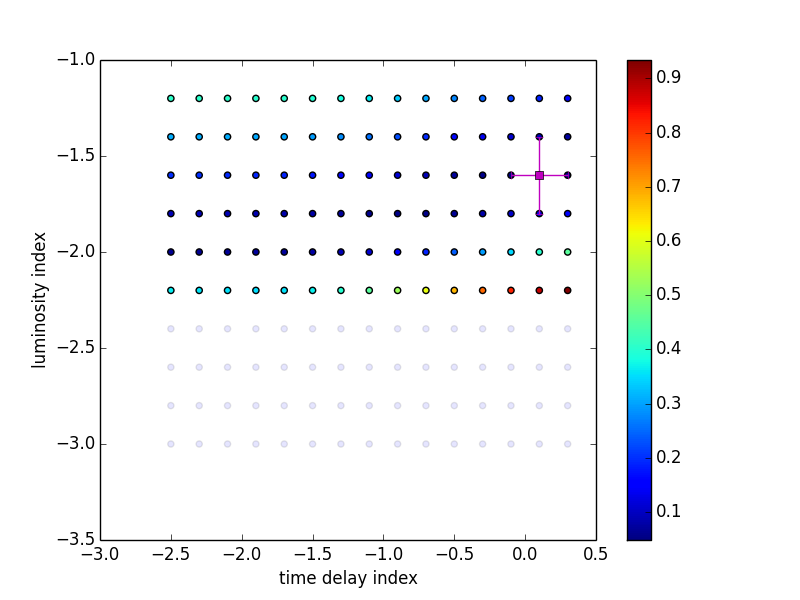}
\caption{Same as figure-3, but for the PL model.}
\end{figure*}
%
%
\begin{figure*} 
\begin{center}
\includegraphics[scale=0.4]{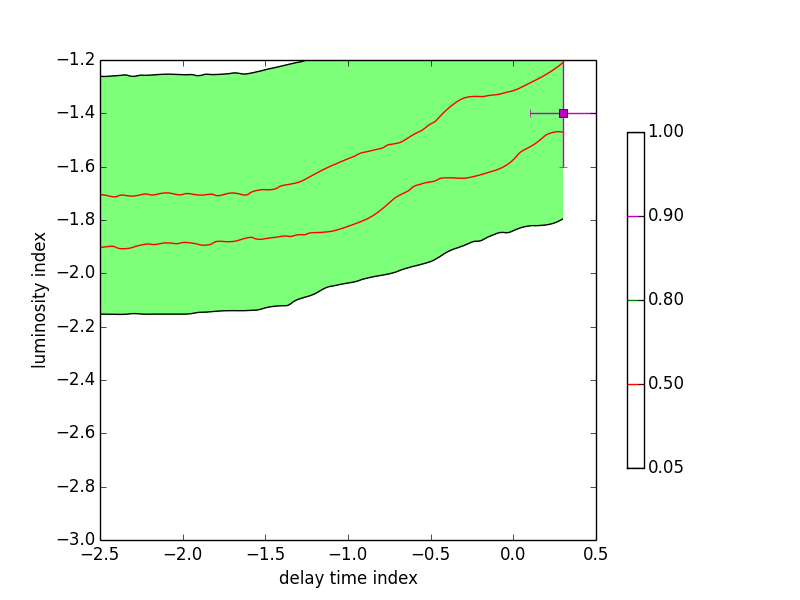}
\includegraphics[scale=0.4]{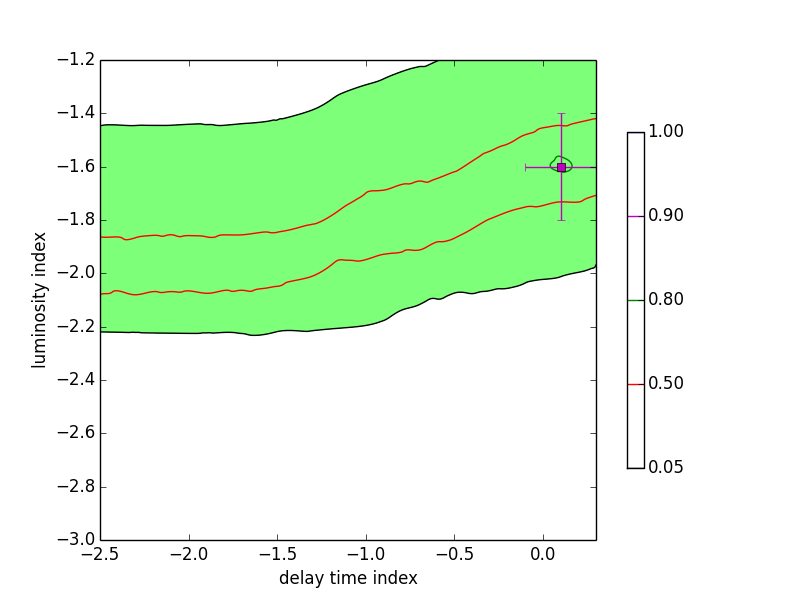}
\caption{Comparison of the KS probability contours and $\chi^2$ grid points for the PL model.}
\end{center}
\end{figure*}
\begin{figure*} 
\begin{center}
\includegraphics[scale=0.5]{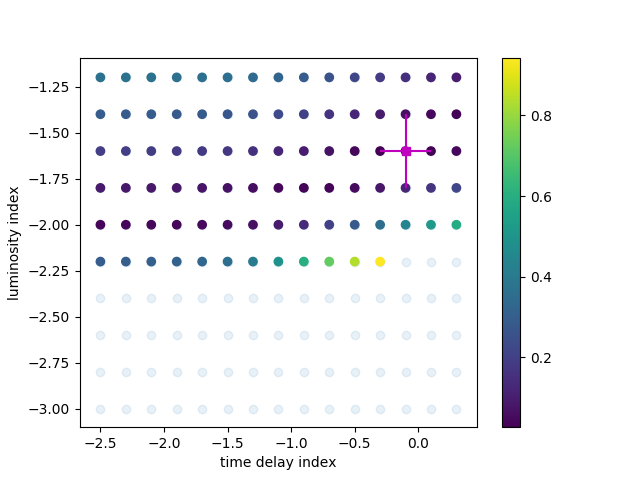}
\includegraphics[scale=0.5]{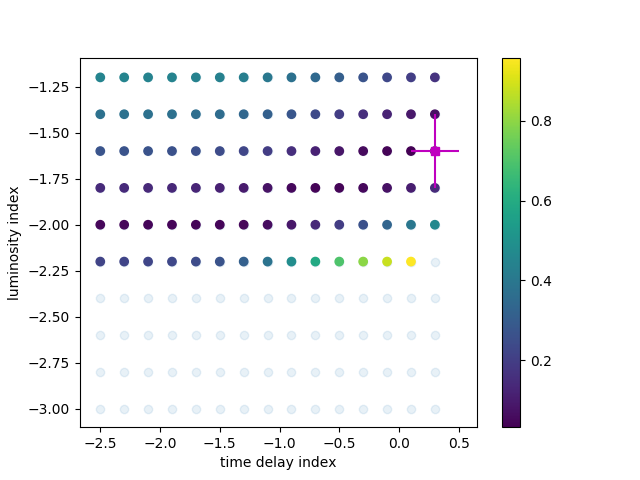}
\caption{Results of the PL model run after including 160410A. The full EE sample shows a shift in the lowest $\chi^2$ (best-fit) point. The shift towards the lower $\eta$ values are within the expectations of the model (see text), and could indicate that any difference in the merger delay time distribution of the EE sample w.r.t to the non-EE sample could merely be a selection effect.}
\end{center}
\end{figure*}
\begin{figure*}
\begin{center}
\includegraphics[scale=0.4]{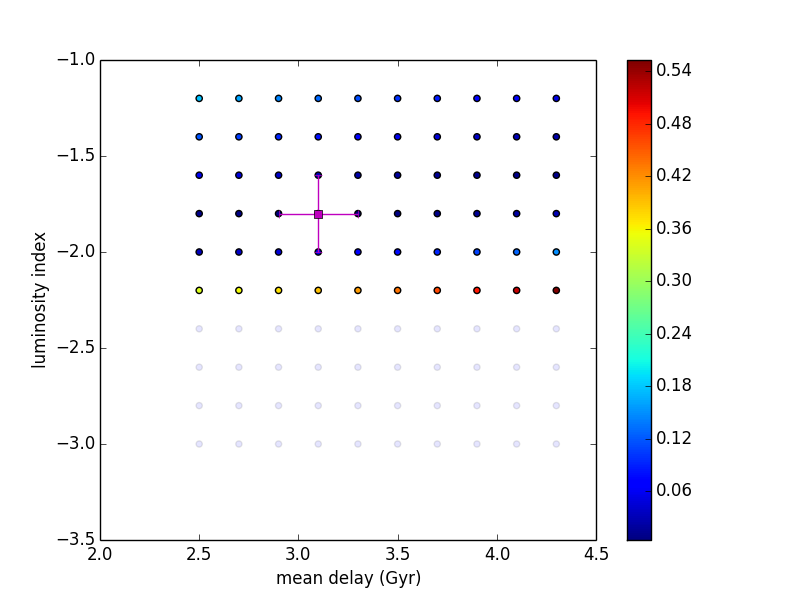}
\includegraphics[scale=0.4]{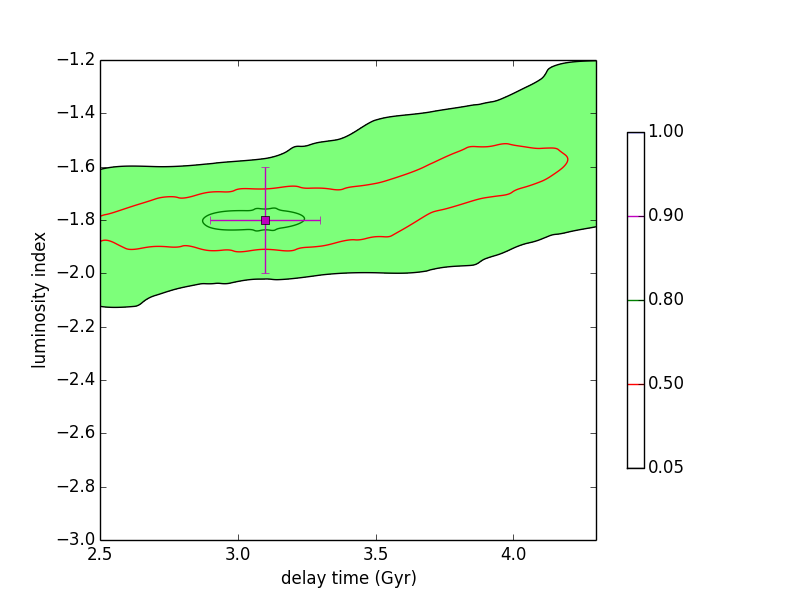}
\caption{As a comparison, this figure shows the non-EE sample with the LN delay time model parameters. The best fit parameters shift to lower delay times compared to the EE sample, but the statistical significance of this difference is limited. Moreover, the somewhat higher delay time of the EE sample could also be due to the faintness that restricts it to lower redshifts.}
\end{center}
\end{figure*}
\begin{figure*}
\begin{center}
\includegraphics[scale=0.4]{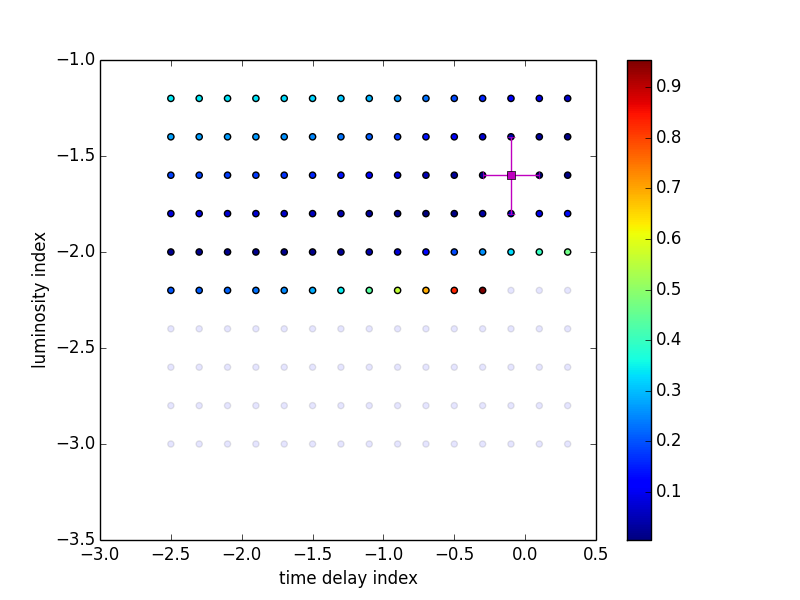}
\includegraphics[scale=0.4]{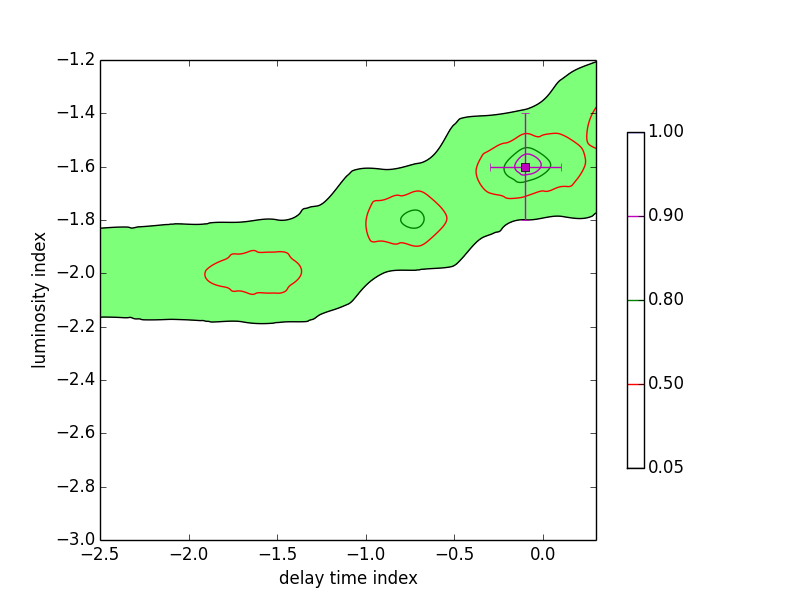}
\caption{$\chi^2$ values on grids and KS test contours for the PL delay time model for the non-EE sample. The colored region is where $\chi^2 \sim DOF$.}
\end{center}
\end{figure*}

\section*{Acknowledgements}
We thank Nial Tanvir, Dipankar Bhattacharya, Paul T O'brien, Sujit Ghosh, and Philp Podsiadlowski for helpful discussions and M. Govindankutty for generously providing us with computing facilities.




\bibliographystyle{mnras}
\bibliography{ref} 




%
%


\bsp	
\label{lastpage}
\end{document}